\pdfoutput=1

\documentclass{article}
\usepackage{ijcai20}
\usepackage{times}
\usepackage{soul}
\usepackage{url}
\usepackage{amsmath}
\usepackage{amsthm}
\usepackage{mathbbol}
\usepackage{txfonts}
\usepackage{booktabs}
\usepackage{makecell}
\usepackage{algorithmic}
\usepackage{wrapfig}
\usepackage[linesnumbered,boxed,ruled]{algorithm2e}
\usepackage{booktabs}
\usepackage{multirow}
\usepackage{stackengine}

\usepackage[colorinlistoftodos]{todonotes}

\newtheorem{theorem}{Theorem}

\pdfoutput=1
\title{Rigorous Explanation of Inference on Probabilistic Graphical Models}

\author{
Yifei Liu$^1$
\and
Chao Chen$^2$\and
Xi Zhang$^1$\And
Sihong Xie$^2$
\affiliations
$^1$Key Laboratory of Trustworthy Distributed Computing and Service\\
Ministry of Education, BUPT,
Beijing, China\\
$^2$Lehigh University,
Bethlehem, PA, USA\\
\emails
liuyifei@bupt.edu.cn,
chc517@lehigh.edu,
zhangx@bupt.edu.cn,
six316@lehigh.edu
}

\begin{document}
\pdfoutput=1
\maketitle
\pdfoutput=1
\begin{abstract}
Probabilistic graphical models, such as Markov random fields (MRF),
exploit dependencies among random variables to model a rich family of joint probability distributions.
Sophisticated inference algorithms, such as belief propagation (BP),
can effectively compute the marginal posteriors.
Nonetheless,
it is still difficult to interpret the inference outcomes for important human decision making.
There is no existing method to rigorously attribute the inference outcomes to the contributing factors of the graphical models.
Shapley values provide an axiomatic framework, but naively computing or even approximating
the values on general graphical models is challenging and less studied.
We propose GraphShapley to integrate
the decomposability of Shapley values,
the structure of MRFs,
and the iterative nature of BP inference
in a principled way for fast Shapley value computation, that
1) systematically enumerates the important contributions to the Shapley values of the explaining variables without duplicate;
2) incrementally compute the contributions without starting from scratches.
We theoretically characterize GraphShapley regarding independence, equal contribution, and additivity.
On nine graphs,
we demonstrate that GraphShapley provides sensible and practical explanations.
\end{abstract}

\section{Introduction}
Probabilistic graphical models (PGMs) play an important role in many real-world applications where dependencies between entities are essential to describe uncertain and complex interactions and dynamics.
For example,
frauds in review and auction networks~\cite{rayana2015collective,Pandit2007},
and potential interests of users on social networks~\cite{namata2016collective}
can be modeled by PGMs and detected by various inference algorithms.
However, the lack of explanations of the models and inferences limits the practical utility.
For example, compared with ``the detection'' of fraudulent accounts,
it is equally, if not more, important to explain ``why'' the detected accounts are suspicious so the end-users can opt to be convinced by the detection,
or to rectify the model for more sensible inferences~\cite{Ross2017}.

\begin{figure}
    \centering
    \includegraphics[width=0.37\textwidth]{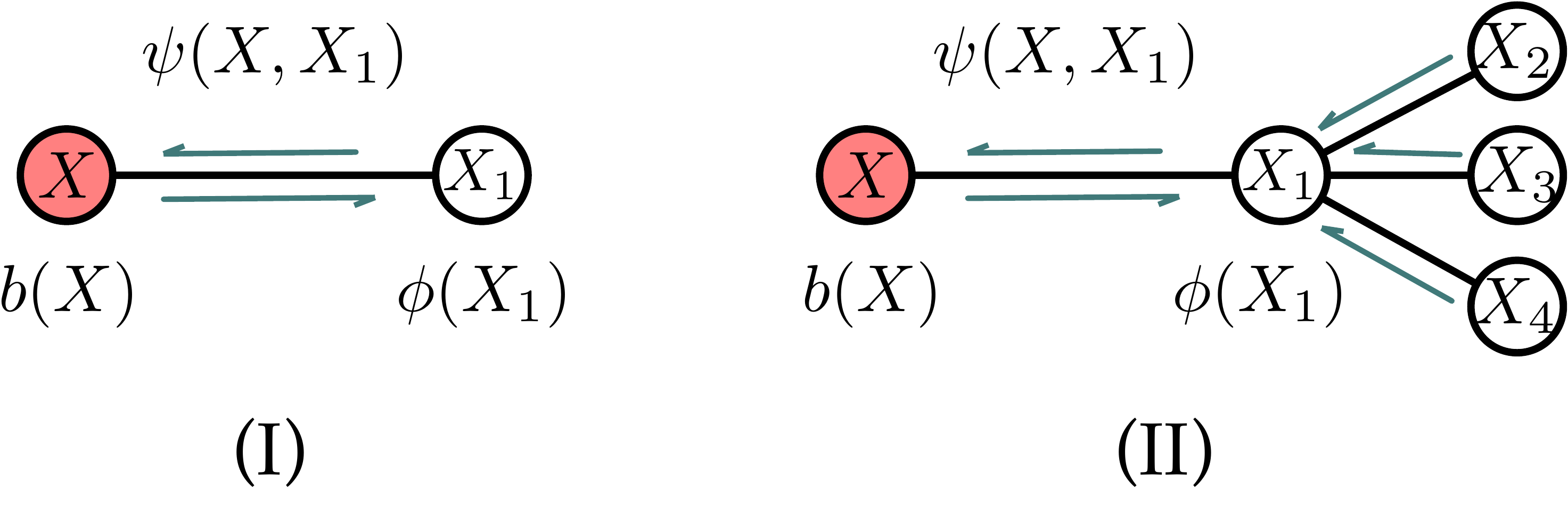}
    \caption{\footnotesize (I): the \textit{probabilistic} contribution of the explaining variable $X_1$ to the target $X$'s belief $b(X)$ depends the prior distribution $\phi(X_1)$ and the dependencies $\psi(X,X_1)$.
    (II): the contribution of $X_1$ to $b(X)$ is compounded with the fact that the contributions from $X_2$, $X_3$, and $X_4$ have to go through $X_1$.
    The Shapley value of $X_1$ should reflect both probabilistic and topological contributions of $X_1$.
    }
    \label{fig:challenges}
\end{figure}
We focus on explaining the results produced by belief propagation (BP) on Markov Random Fields (MRF).
As shown in Figure~\ref{fig:challenges},
an MRF describes the dependencies among random variables (shown as circles).
Each variable is assigned a prior distribution (e.g., $\phi(X)$), representing prior knowledge without the dependencies.
The dependencies among the variables are 
represented by the edges, each of which has compatibility parameters $\psi(X,X_1))$ that describe how one variable can influence its neighbors.
BP computes the beliefs (posteriors) of the random variables (e.g., $b(X)$),
by passing messages (the arrows over the edges)
between variables until convergence.
When the beliefs are used to make critical decisions, we aim to find the top salient factors
that contribute to the belief of the target variables (e.g., those representing potential spammer accounts)
to aid human decision makings.

The dependencies are a double-edged sword:
on the one hand, the dependencies more accurately capture the interactions between the variables when a variable does depend on other variables 
(e.g., in spam detection, an account is more suspicious
if it posts suspicious reviews to dishonest businesses);
on the other hand, 
when inferring the belief of a target variable (such as to test if an account is a spammer),
the dependencies compound with many factors so that the inferences cannot be straightforwardly interpreted by the decision-makers.
For example, an account is deemed suspicious since it posted reviews to some dishonest businesses,
whose beliefs can further depend on thousands of reviews.

Existing explanation algorithms that have been devised for Explainable AI are not sufficient (see the excellent surveys of~\cite{Du2019,Guidotti2018}).
The methods in
\cite{Ribeiro2016LIME,Shrikumar2017DEEPLIFT} assign an importance score to each feature without taking into account the dependencies between the features.
On graphs, the authors of
\cite{ying2019gnn} explain arbitrary graph neural networks using gradients but do not consider PGMs.
The methods proposed in~\cite{darwiche2003differential,chan2005sensitivity,chen2019scalable} are more relevant to PGM inference explanations: they use gradients or greedy search to find salient factors to explain the inference.
Unlike Shapley values,
these methods lack a rigorous characterization of the attribution.

On explainable MRF inferences,
we argue that a more principled explanations
is necessary to complement the prior work~\cite{darwiche2003differential,chan2005sensitivity,chen2019scalable}.
We will adopt Shapley's framework,
which provides a fair attribution of a total gain to the players in a cooperative game~\cite{shapley1953value}.
When used to explain a machine learning model,
the explaining variables are the players
and the game is the output of the model.
Since an MRF represents a probability distribution using a graph of random variables, the desired explanations should reflect the importance of an explaining variable based on its probabilistic and topological properties.
See Figure~\ref{fig:challenges} for examples.
In~\cite{Shrikumar2017DEEPLIFT},
multiple explanation methods for classification are unified as Shapley value computation.
In~\cite{Skibski2019}, they dealt with Shapley and Myerson values for graph-restricted games rather than PGMs.
In~\cite{chen2018L-shapley}, only linear chains and grids are considered. 
None of these methods quantify the probabilistic and topological importances of the random variables.

We propose GraphShapley,
to compute a Shapley value SV$(X_i; X, G)$ that
measures the topological and probabilistic contribution of any explaining variable $X_i$
to the belief $b(X)$ of a target random variable $X$ on a given MRF $G$.
GraphShapley enumerates all possible subgraphs where $X_i$ and $X$ are connected.
We average the contribution of $X_i$ to $b(X)$ on each subgraph,
which is measured by the change in $b(X)$ before and after $X_i$ is omitted.
For example, in Figure~\ref{fig:challenges} (II),
by removing $X_1$,
the variables $X_2,X_3,X_4$
can not contribute to $b(X)$,
which can change dramatically,
and we can learn about the topological importance of $X_1$ on this subgraph.
This topological contribution cannot be measured
by alternating $X_1$'s 
prior~\cite{chen2018L-shapley} only,
where $X_1$ and its other connections still play a role in message passing.
These subgraphs also help measure the probabilistic importance of $X_i$ since the prior $\phi(X_i)$ will be removed as $X_i$ is omitted (try removing $X_1$ from the subgraph in Figure~\ref{fig:challenges} (I)).

\textbf{Conceptually}, we provide a novel definition of Shapley values to quantify the topological and probabilistic contribution of random variables on an MRF.
\textbf{Computationally}, enumerating all such subgraphs can be expensive and we propose to approximate $\textnormal{SV}(X_i; X, G)$ on a small neighborhood of the target $X$ that contains most of the salient explaining variables.
We design a depth-first subgraph search (Algorithm~\ref{alg:GraphShapleyAlg}) that avoids duplicated enumeration while including all relevant ones.
By exploiting the iterative computations in BP, 
we retrieve the cached results from one subgraph to incrementally compute the messages on larger subgraphs during the DFS search.
\textbf{Theoretically}, we prove several properties that GraphShapley has to deliver deeper insights.
\textbf{Empirically}, on nine MRFs, we conduct experiments to confirm that GraphShapley can better identify influential explaining variables,
compare to the gradient-based and other explanation methods.
\section{Problem Formulation}
An MRF is a graph $G=({\cal V},{\cal E})$, where ${\cal V}$ is the set of random variables.
As commonly found in multi-class classification problems~\cite{rayana2015collective},
we assume that each random variable $X_i$ is discrete,
taking values $X_i=x_i$ from $c$ classes $[c]=\{1,\dots, c\}$.
Each random variable $X_i\in{\cal V}$ has a prior distribution $\phi(x_i)\in\mathbb{R}^{c}$,
and each edge $(X_i, X_j)\in \mathbb{R}^{c\times c}$ has 
the compatibility matrix $\psi(x_i,x_j)$
to encode the likelihood of $X_i$ and $X_j$ taking value $(x_i, x_j)$ jointly. 
The graph $G$ factorizes the joint distribution $P({\cal V})$ as
\begin{equation}
\label{eq:joint_dis}
P({\cal V}) = \frac{1}{Z} \sum_{X_i\in {\cal V}} \phi(X_i)\prod_{X_j\in {\cal N}(X_i)} \psi(X_i, X_j),
\end{equation}
where $Z$ normalizes the product to a probability distribution,
and $\mathcal{N}(X_i)=\{X_j|(X_j, X_i)\in {\cal E}\}$ is the neighbors of $X_i$.

Computing the marginal distributions of the variables can be expensive since there are exponentially many cases to marginalize. 
Belief propagation (BP) is an approximation algorithm to compute the marginal distributions $b(X)$ by message passing.
Specifically,
$m_{j\to i}(x_i) $ is the message from $X_j$ to $X_i$, formally defined as:
\begin{equation}
\label{eq:sum_prod}
\frac{1}{Z_i}\sum_{x_j\in [c]}
\left[\psi(x_i, x_j)\phi(x_j) \prod_{
k\in {\cal N}(X_j)\setminus \{i\}
} m_{k\to j}(x_j)\right],
\end{equation}
where $Z_i$ is a normalization factor.
The belief (marginal posterior) of $X_i$,
denoted by $b(X_i)$,
can be inferred using:
\begin{equation}
\label{eq:belief}
    b(X_i) \propto \phi(X_i)\prod_{X_j\in {\cal N}(X_i)} m_{j\to i}(X_i).
\end{equation}

It can be seen that $b(X_i)$ is not only related to $\phi(X_i)$, but also related to the incoming messages that recursively depend on other priors, edge potentials, and messages.
This makes graph inference less transparent to a human end-users and calls for explanations of the inference process or outcomes by
providing the \textit{causes} of the inference outcomes.
Prior MRF explanation methods~\cite{chen2019scalable,chan2005sensitivity} focused on explaining the internal computations of the inference algorithm, while we propose another form of explanations
to help the end-users discover the important factors that lead to the inference outcomes,
without explaining \textit{how} the belief is computed, which is the responsibilities of the above prior work.

Consider the BP algorithm as an inference game, where all variables (or players) from ${\cal V}$ form a coalition to collectively contribute to the computation of the belief $b(X)$.
We propose to compute Shapley values of the players to fairly attribute $b(X)$ to all variables in the coalition ${\cal V}$.
The top few variables receiving the most attributions can be regarded as a succinct explanation to 
why the belief $b(X)$ is as such.

The Shapley value of $X_i$ when contributing to $b(X)$ on $G$ is defined as 
the average of $X_i$'s contributions to $b(X)$ in all possible subgraphs (or coalitions) $S\subset G$ that contains $X$ and $X_i$, denoted by ${\cal S}(X_i;X,G)$.
Each subgraph/coalition can be used as a new MRF where BP can compute a belief $\tilde{b}(X)$ to approximate $b(X)$.
We define the characteristic function $\nu:{\cal S}\to \mathbb{R}$ to evaluate
the quality of a coalition $S\in {\cal S}$
in approximating $b(X)$.
As KL-divergence is a well-established measurement of approximating a probability distribution~\cite{chen2019scalable,Suermondt1992},
we adopt the following symmetric KL-divergence between the two beliefs $b(X)$ and $\tilde{b}(X)$:
\begin{align}
\label{eq:characteristic_function}
\nu(S;X) &= -\textnormal{KL}(b(X)||\tilde{b}(X)) - \textnormal{KL}(\tilde{b}(X)||b(X)) \\\nonumber
&= -\sum_x b(x) \log[b(x) /\tilde{b}(x)]
- \sum_x \tilde{b}(x) \log[\tilde{b}(x) / b(x)]\\\nonumber
&=H(b)+H(\tilde{b})-\textnormal{NLL}(b,\tilde{b})-\textnormal{NLL}(\tilde{b}, b),
\end{align}
where $H(b)=-\sum_x b(x)\log b(x)$ is the entropy of the distribution $b(X)$
and $\textnormal{NLL}(b, \tilde{b})=-\sum_x b(x)\log \tilde{b}(x)$ is the negative log-likelihood loss when using $\tilde{b}(X)$ to predict $b(X)$, and likewise for $H(\tilde{b})$ and NLL$(\tilde{b}, b)$.
A higher $\nu$ indicates that $\tilde{b}(X)$ can approximate $b(X)$ well without over-committing to a particular class,
similar to the maximum entropy classifier~\cite{McCallum00}.
The characteristic function in~\cite{chen2018L-shapley} is just $-$NLL$(b, \tilde{b})$,
which is not symmetric and is a lower-bound of $-\textnormal{KL}(b(X)||\tilde{b}(X))$.
We directly evaluate the approximation quality of $S$ without further resorting to the lower-bound.

We define the \textit{marginal contribution} of $X_i$ to $b(X)$ when $X_i$ works within the coalition $S$,
as the difference in the approximation quality with and without $X_i$:
\begin{equation}
\label{eq:mx}
\mu(X_i; X, S) = \nu(S, X) - \nu(S \setminus \{X_i\}, X).
\end{equation}

The Shapley value of $X_i$ when contributing to $b(X)$ on $G$ is then
obtained by averaging the marginal contributions over all coalitions in ${\cal S}(X_i; X, G)$:

\begin{equation}
\label{eq:sv}
\textnormal{SV}(X_i; X, G) :=  \frac{1}{|{\cal S}(X_i; X, G)|} \sum_{S\in {\cal S}(X_i; X, G)}\mu(X_i; X, S)
\end{equation}

Note that Eq. (\ref{eq:sv}) is not an approximation but the exact definition of the Shapley values.
The steps for evaluating $\textnormal{SV}(X_1; X, G)$ is shown in Figure~\ref{fig:graph_shapley} over a simple MRF.
On larger MRFs in real-world applications,
computing $\textnormal{SV}(X_i; X, G)$ is challenging since: 1) ${\cal S}(X_i; X, G)$ is exponentially large, 2) the enumeration all coalitions needs a carefully design search, and 3) the evaluation of the characteristic function $\nu$ on each coalition require running BP to estimate $\tilde{b}(X)$, which is costly.
Prior work~\cite{chen2018L-shapley,Skibski2019} computes Eq. (\ref{eq:sv}) in a combinatoric manner by
enumerating all possible subsets of the random variables.
This is not applicable for MRFs since the same set of variables can be connected in multiple ways, as shown by the coalitions $S_2$ and $S_3$ in Figure~\ref{fig:graph_shapley}.

\section{Method}
\begin{figure}[t]
    \centering
    \includegraphics[width=0.45\textwidth]{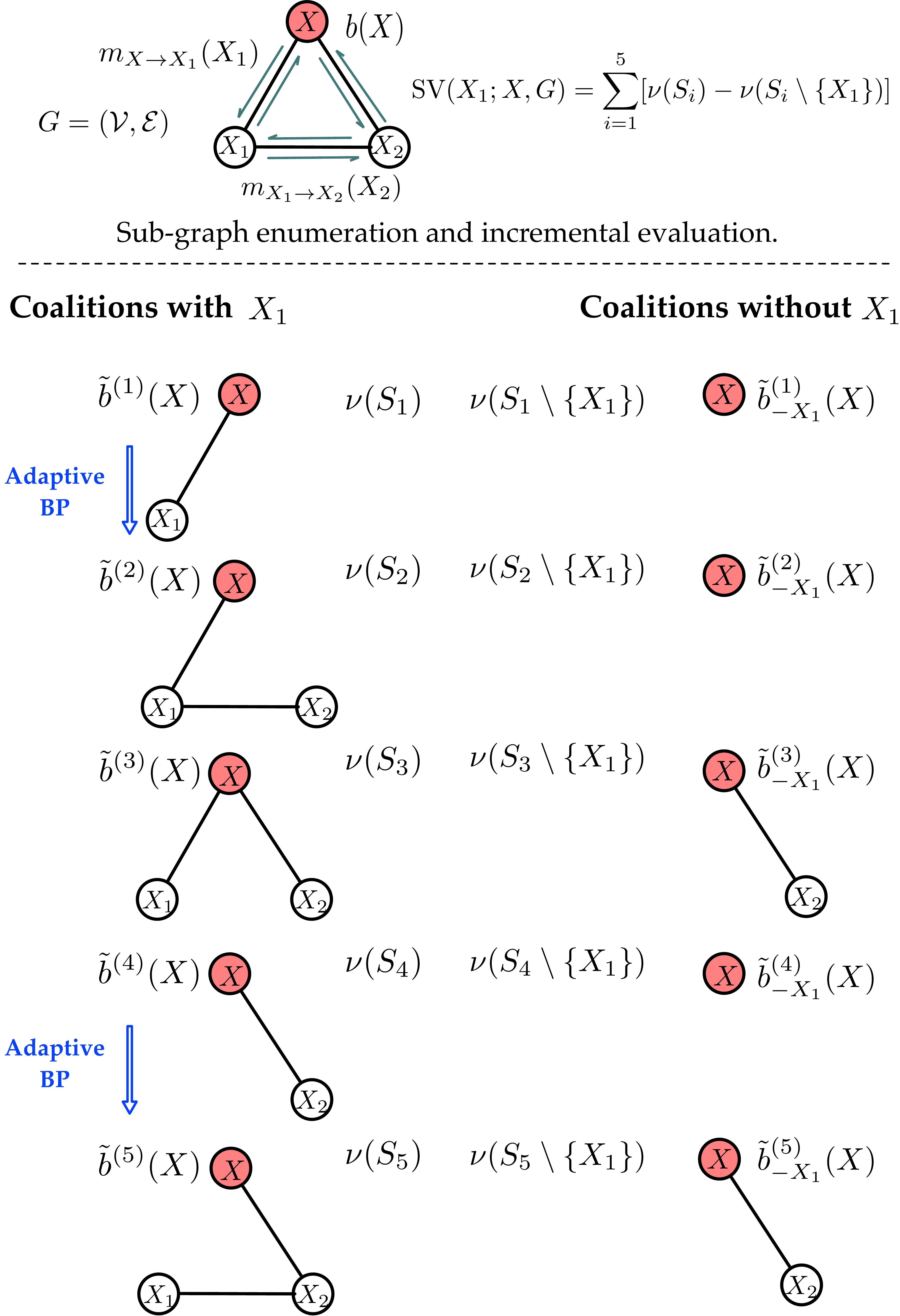}
    \caption{An example of calculating Shapley value SV$(X_1; X, G)$ on $G$ with subgraph enumeration and incremental evaluation.}
    \label{fig:graph_shapley}
\end{figure}

To address the above challenges,
we propose GraphShapley (Algorithm \ref{alg:GraphShapleyAlg}),
for effective and efficient computation of $\textnormal{SV}(X_i; X, G)$.
By exploiting the iterative nature of BP inference on an MRF,
our algorithm considers the followings:
1) restricting the maximum search distance to approximate Shapley value,
taking into account the variables that truly contribute,
2) scheduling search algorithm to avoid duplicated enumeration and to ensure the completeness of the enumerated subgraphs connected in multiple ways, and
3) re-using existing outcomes to incrementally measure the contributions,
reducing the computational cost.

\begin{algorithm}[t!]
\DontPrintSemicolon
\footnotesize
\SetAlgoLined
\caption{GraphShapley}
\label{alg:GraphShapleyAlg}
\SetKwData{Left}{left}
\SetKwData{This}{this}
\SetKwData{Up}{up}
\SetKwFunction{Union}{Union}
\SetKwFunction{FindCompress}{FindCompress}
\SetKwInOut{Input}{Input}
\SetKwInOut{Output}{Output}
\Input{Graph $G=({\cal V},{\cal E})$, a target variable $X \in {\cal V}$ to be explained with its belief $b(X)$, maximum search distance $D$ from $X$ to $X_i$, maximum subgraph complexity $C$}
\Output{Shapley Value SV$(X_i;X,G)$ for each explaining variable $X_i$}
\SetKwFunction{FMain}{GraphShapley}
    \SetKwProg{Fn}{Function}{:}{}
    \Fn{\FMain{}}{
    Sort $\mathcal{N}(X_i)$ by degree for each $X_i \in G$ \;
    DFSEnumerate$(G,(X),X,\emptyset)$ \;   
    SV$(X_i; X, G) = \frac{1}{|{\cal S}(X_i; X, G)|} \sum_{S\in {\cal S}(X_i; X, G)}\mu(X_i; X, S)$ \; 
    }
    \textbf{End Function}\;
 \BlankLine
    \SetKwFunction{FMain}{DFSEnumerate}
    \SetKwProg{Fn}{Function}{:}{}
    \Fn{\FMain{G, Sub, v, Forbidden}}{
    // $Sub$:current subgraph;  $v$:current node to explore \; 
     \ForEach{u $\in \mathcal{N}(v)$}{
        \uIf{$(v,u) \not\in$ Fobidden $\land$ u $\not\in$ Sub $\land$  // keep acyclic   \\ 
        $len(Sub)<C$ $\land$ $d(u,x)<D$   
        }
            {IncrementalEvaluation$(Sub,(v,u))$  \;     
            DFSEnumerate($G,Sub\cup\{(v,u)\},u,Forbidden$)\;    
            $Forbidden \longleftarrow Forbidden \cup\{(v,u)\}$ \; }
        }
        \ForEach{m $\in$ $Sub \setminus \{v\}$}{ // \textit{expand from other variables}
        DFSEnumerate($G,Sub,m,Forbidden$)\;
        }
    }
    \textbf{End Function}\;
    \SetKwFunction{Fbp}{IncrementalEvaluation}            
    \SetKwProg{Fnn}{Function}{:}{}
    \Fn{\Fbp{S, (v, u), X}}{
        $\tilde{b}(X)=$AdaptiveBP$(Sub,(v,u),X)$ \;
        \ForEach{$X_i \in S$}{
        \uIf{$S \setminus \{X_i\} $ not visited}
        {Find the largest subgrpah of $S \setminus \{X_i\}$ that has BP results.\\
        Start from that subgraph and use Adaptive BP to find $\tilde{b}_{-X_{i}}(X)$ for $X_i \in S$.\\
        }
        Evaluate and record $\mu(X_i; X, S) = \nu(S, X) - \nu(S \setminus \{X_i\}, X)$. \;
        }
    }
    \textbf{End Function}\;
    \SetKwFunction{Fbp}{AdaptiveBP}            
    \SetKwProg{Fnn}{Function}{:}{}
    \Fn{\Fbp{S, (v, u), X}}{
      // $S$:current MRF with converged messages;\\
      // $v$: variable in $S$; $u$: a new variable to be added to $S$\\
        $S\leftarrow S\cup (v,u)$\\
        Use Adaptive BP to compute the new belief $\tilde{b}(X)$.\\
        \textbf{Return} $\tilde{b}(X)$ 
    }
    \textbf{End Function}
\end{algorithm}

\noindent \textbf{DFS Subgraph Enumeration}
Depth-first search makes full use of the topology of graphical models to explore all possible connected subgraphs recursively.
Due to the expensive cost of enumerating all subgraphs on a large scale graph,
we consider maximum search distance $D$ from the target variable,
beyond which the variables will not be involved,
where $d(a,b)$ is used to obtain the shortest path distance from $a$ to $b$,
and maximum subgraph complexity $C$ to limit the size of the enumerated subgraph.
A divide-and-conquer technique is applied to enumerate acyclic subgraphs connected in multiple ways,
where edges are considered instead of nodes.
The enumeration is divided into two parts:
in the first one,
starting from the target variable,
we explore the subgraphs containing further edges through depth-first search;
in the second one,
we expand the subgraphs from other variables on the subgraph recursively.
Forbidden edges are used to record the edges,
which DFS has completed and will not visit in the future search.
A newly explored edge will not be added to the subgraph \textit{Sub} if:
1) the edge has already in forbidden edges, or
2) its addition will lead to a cycle, or
3) its addition will make the subgraph larger than the capacity $C$, or
4) the new node is $D$ hops away.
We do not process another edge until the edge from the previous one is fully-processed.
The enumeration of subgraphs will not be completed until all edges have been processed.

We briefly analyze the example in Figure \ref{fig:graph_shapley} of calculating Shapley value SV$(X_1; X, G)$ on a small graph $G$.
Starting from the subgraph containing only the target variable $\{X\}$,
$S_1$ and $S_2$ are obtained through adding $X_1$ and $X_2$ respectively (line 12).
Since the addition of edge $(X_2,X)$ to $S_2$ will lead to a cycle (line 9),
$S_2$ is back-tracked to $S_1$ and $(X_1,X_2)$ is added to forbidden edges (line 13).
$X_1$ in $S_1$ has been explored and $S_3$ is obtained by extending $X$ in $S_1$ (line 16).
Exploration can not proceed from $S_3$,
so the algorithm returns from the recursive call at line 11 and adds $(X,X_1)$ to forbidden edges (line 13).
By getting to $X_2$ from $X$, $S_4$ and $S_5$ will be obtained in order.
Note that the same set of variables can be connected in multiple ways, resulting in different coalitions/subgraphs,
such as $S_2$ and $S_5$ in Figure \ref{fig:graph_shapley}. 
Perform a rollback and complete the search after $S_5$.

\noindent\textbf{Adaptive Belief Propagation}
The marginal contributions of explaining variables need to be measured over all enumerated subgraphs,
and usually each subgraph is modeled as an MRF separately where BP infers marginal distribution,
regardless of the connections between the graphs.
We consider \textit{adaptive belief propagation}~\cite{Papachristoudis2015adaptivebp}
that recycles the converged messages on a smaller MRF ($Sub$ in line 10 of Algorithm~\ref{alg:GraphShapleyAlg})
to speed up the convergence of the message computations on a larger MRF ($Sub\cup (v,u)$ in line 28),
thus avoiding running BP from scratches whenever a new edge is added.
For example, considering two MRFs
$S_1$ and $S_2$ in Figure \ref{fig:graph_shapley} that $S_2$ is visited by DSFEnumerate right after $S_1$,
the belief $b^{(2)}(X)$ on $S_2$ can be computed using the converged messages on $S_2$, starting from the already converged messages on $S_1$.
Prior~\cite{Papachristoudis2015adaptivebp} and our experiments show that Adaptive BP is faster.

\noindent\textbf{Shapley Values Calculation}
The calculation of Shapley value is accompanied by subgraph enumeration,
while the contributions of the variables will be incrementally measured once a subgraph is enumerated.
Shapley values SV$(X_i;X,G)$ for each explaining variable $X_i$ for the target variable $X$ will be obtained (line 4),
when the subgraph enumeration on $G$ is completed.
For example,
SV$(X_1;X,G)$ is the average marginal contribution of $X_1$ to $X$ over all enumerated subgraphs (line 4),
and SV$(X_2;X,G)$ can be obtained together with SV$(X_1;X,G)$).

\section{Theoretical Analysis of GraphShapley}
We first confirm that GraphShapley conforms with the independence between the target and explaining variables.
\begin{theorem}
\textnormal{(\textbf{Independence})}
If $X$ and $X_i$ are disconnected in $G$ so that $X\Perp X_i$,
then SV$(X_i;X,G)=0$. 
Further, if $X_i$ is connected to $X$
but blocked by the Markov Blanket $B\subset {\cal V}$ of $X$ so that $X\Perp X_i | B$,
then SV$(X_i;X,G)=0$.
\end{theorem}
The first statement in the theorem is obvious based on the definition of SV$(X_i;X,G)$ and Algorithm~\ref{alg:GraphShapleyAlg}.
The second statement can be proved using the definition of SV$(X_i;X,G)$ and the definition of Markov Blanket~\cite{koller2009pgm}.
As an example, in Figure~\ref{fig:challenges} (II), if $X_1$ is in the Markov blanket of $X$, then $X_2$ will have zero contribution of $b(X)$. On the other hand, in the MRF $G$ in Figure~\ref{fig:graph_shapley}, even if $X_1$ is in the Markov blanket of $X$, $X_2$ can still contribute to $b(X)$ through the path $(X, X_2)$.

\begin{theorem}
\label{thm:equal}
\textnormal{(\textbf{Equal contribution})}
Given any two variables $X_i$ and $X_j$,
if $\nu(S\cup \{X_i\};X)=\nu(S\cup \{X_j\};X)$ for any coalition $S$,
then \textnormal{SV}$(X_i;X;G)=$\textnormal{SV}$(X_j;X;G)$.
\end{theorem}

\begin{theorem}
\textnormal{(\textbf{No Additivity})}
There exists an MRF $G=({\cal V},{\cal E})$ and a random variable $X\in{\cal V}$, such that
$\sum_{i}\textnormal{SV}(X_i;X,G)\neq \nu({\cal V})-\nu(\emptyset)$
\end{theorem}
The lack of additivity is due to the removal of random variables and the associated edges.
Additivity is satisfied by the Shapley values in~\cite{chen2018L-shapley},
as they don't alter the MRF topology.
It is future work to show whether additivity and topological importance measurement are compatible. 


\section{Experiments}

\vspace{.05in}
\noindent\textbf{Datasets and MRF setups}
\vspace{.05in}

\noindent We drew datasets from three applications. 
The statistics of the datasets are shown in Table \ref{tab:overall} (left panel).
First, in collective classification,
we construct an MRF for each of the three citation networks (Citeseer, Cora, PubMed),
with the research area that a paper belongs to as a random variable (node),
and a paper citation as an undirected edge \cite{namata2016collective}.
We assume a homophily relationship over the edges,
meaning two papers are likely to be in the same area if one cites the other.
The beliefs inferred by BP are the posterior class distributions of the papers.
We randomly select 80\% of the variables with class-biased priors and the remaining 20\% have uniform priors.
The explanations are computed on the 20\% portion.
Second, we adopt the Yelp review networks for spam detection.
We represent reviewers, reviews, and products and their relationships by an MRF and set node priors and compatibility matrix following the state-of-the-art MRF-based spam detector proposed in~\cite{rayana2015collective}.
There 
Lastly, we represent users (Blogcatalog, Flickr and Youtube) as nodes and behaviors including subscription and tagging as edges \cite{tang2009relational}.
BP infers the preferences of users. 
The experimental setups are the same as the citation networks.

\begin{table*}[!htb]
\centering
\scriptsize


\begin{tabular}{c
                ||@{\hspace*{1mm}}c
                |@{\hspace*{1mm}}c
                |@{\hspace*{1mm}}c
                |@{\hspace*{1mm}}c
                ||c
                  @{\hspace*{1mm}}c
                  @{\hspace*{1mm}}c 
                  @{\hspace*{1mm}}c 
                  @{\hspace*{1mm}}c 
                  @{\hspace*{1mm}}c 
                  @{\hspace*{1mm}}c}
    \multirow{2}{*}{\textbf{Datasets}} &
    \multicolumn{4}{c||}{\textbf{Data Property}} &
    \multicolumn{7}{c}{\textbf{Performance}} \\
 \cline{2-12}

& 
\addstackgap[2.8pt]{\textbf{Classes}} &
\textbf{Nodes} &
\textbf{Edges} &
\textbf{edge/node} &
\textbf{Random} & 
\textbf{Embedding} & 
\textbf{PageRank} & 
\textbf{Sensitivity} & 
\textbf{LIME} & 
\textbf{MC-sampling} & 
\textbf{GraphShapley}
\\

\hline
Cora
& \makecell[c]{7}
& \makecell[c]{2,708}
& \makecell[c]{10,556}
& \makecell[c]{3.90}
& \makecell[c]{0.831}
& \makecell[c]{0.344} 
& \makecell[c]{0.891}
& \makecell[c]{0.729} 
& \makecell[c]{1.401} 
& \makecell[c]{0.218 $\pm$ 0.03 $\circ$}
& \makecell[c]{\textbf{0.119} $\bullet$}

\\

Citeseer
& \makecell[c]{6}
& \makecell[c]{3,321}
& \makecell[c]{9,196}
& \makecell[c]{2.78}
& \makecell[c]{0.495}
& \makecell[c]{0.301} 
& \makecell[c]{0.589}
& \makecell[c]{0.512} 
& \makecell[c]{0.921} 
& \makecell[c]{0.179 $\pm$ 0.03 $\circ$}
& \makecell[c]{\textbf{0.078} $\bullet$}
\\

PubMed
& \makecell[c]{3}
& \makecell[c]{19,717}
& \makecell[c]{44,324}
& \makecell[c]{2.25}
& \makecell[c]{1.043}
& \makecell[c]{0.706} 
& \makecell[c]{1.118}
& \makecell[c]{0.941} 
& \makecell[c]{1.519} 
& \makecell[c]{0.431 $\pm$ 0.14 $\circ$}
& \makecell[c]{\textbf{0.092} $\bullet$}
\\


YelpChi
& \makecell[c]{2}
& \makecell[c]{105,659}
& \makecell[c]{269,580}
& \makecell[c]{2.55}
& \makecell[c]{0.296}
& \makecell[c]{0.058} 
& \makecell[c]{0.035}
& \makecell[c]{0.011 $\circ$ }
& \makecell[c]{0.691} 
& \makecell[c]{0.038 $\pm$ 0.01 }
& \makecell[c]{\textbf{0.001} $\bullet$}
\\

YelpNYC
& \makecell[c]{2}
& \makecell[c]{520,200}
& \makecell[c]{1,436,208}
& \makecell[c]{2.76}
& \makecell[c]{0.297}
& \makecell[c]{0.058} 
& \makecell[c]{0.043}
& \makecell[c]{0.018 $\circ$} 
& \makecell[c]{0.692} 
& \makecell[c]{0.042 $\pm$ 0.01}
& \makecell[c]{\textbf{0.001} $\bullet$}
\\

YelpZip
& \makecell[c]{2}
& \makecell[c]{873,919}
& \makecell[c]{2,434,392}
& \makecell[c]{2.79}
& \makecell[c]{0.204}
& \makecell[c]{0.084} 
& \makecell[c]{0.031}
& \makecell[c]{0.012 $\circ$}
& \makecell[c]{0.693} 
& \makecell[c]{0.027 $\pm$ 0.01}
& \makecell[c]{\textbf{0.001} $\bullet$}
\\

Blogcatalog
& \makecell[c]{39}
& \makecell[c]{10,312}
& \makecell[c]{333,983}
& \makecell[c]{32.39}
& \makecell[c]{6.673}
& \makecell[c]{6.285} 
& \makecell[c]{3.903}
& \makecell[c]{5.944}
& \makecell[c]{-} 
& \makecell[c]{3.323 $\pm$ 1.17 $\circ$}
& \makecell[c]{\textbf{2.397} $\bullet$}
\\

Flickr
& \makecell[c]{195}
& \makecell[c]{80,513}
& \makecell[c]{5,899,882}
& \makecell[c]{73.28}
& \makecell[c]{3.695}
& \makecell[c]{3.082} 
& \makecell[c]{2.789}
& \makecell[c]{2.650 $\circ$}
& \makecell[c]{-} 
& \makecell[c]{2.833 $\pm$ 0.31}
& \makecell[c]{\textbf{1.622} $\bullet$}
\\

 Youtube
& \makecell[c]{47}
& \makecell[c]{31,703}
& \makecell[c]{96,361}
& \makecell[c]{3.04}
& \makecell[c]{0.077}
& \makecell[c]{0.061} 
& \makecell[c]{0.074}
& \makecell[c]{0.070} 
& \makecell[c]{-} 
& \makecell[c]{0.044 $\pm$ 0.01 $\circ$}
& \makecell[c]{\textbf{0.031} $\bullet$}
\\
\hline
\end{tabular}
\caption{\textbf{Left}: Statistic of Datasets.
\textbf{Right}: Overall symmetric KL performances ($\circ$ indicate the runner-up methods and $\bullet$ indicate significance in $t$-test). LIME does not apply to the last three datasets since there are no node features.}

\label{tab:overall}
\end{table*}

\vspace{.05in}
\noindent\textbf{Baselines}
\vspace{.05in}

\noindent\textbf{Random} generates an importance score of each explaining variable for each target variable randomly, ignoring the messages in BP.

\noindent\textbf{Embedding} utilizes DeepWalk~\cite{Bryan2014deepwalk} to obtain embedded representation of variables based on node proximity on the MRFs,
and variable importance scores are assigned based similarity of explaining nodes to the target node.

\noindent\textbf{PageRank} \cite{page1999pagerank} is a global ranking of the importances of the variables, regardless of the target node to be explained.
It also fails to consider the beliefs and messages generated by BP.

\noindent\textbf{Sensitivity Analysis} \cite{chan2005sensitivity} is a gradient-based approach that measures output changes due to input changes.
We approximate the gradient of the target beliefs with respect to an explaining variable's prior by comparing the beliefs before and after setting the explaining variable's prior to the uniform distribution. 
It fails to consider topological contribution
as the gradients are estimated
using perturbations in the priors rather than the graph topology.

\noindent\textbf{LIME} \cite{Ribeiro2016LIME} fits a logistic regression model on node features
to predict the beliefs on the target variables,
(see \cite{chen2019scalable} for the details).

\noindent\textbf{MC-sampling}
Monte Carlo simulation can approximate Shapley values~\cite{CASTRO20091726,Strumbelj2014}.
We follow Eq. (\ref{eq:sv}) as in GraphShapley but replace the DSF-enumerated subgraphs with randomly sampled spanning trees rooted at the target nodes.

We do not compare GraphShapley with GraphEXP \cite{chen2019scalable}, since the form of explanation is different (a ranking of explaining variables vs. a subgraph for each target).
The evaluations are also different,
we retain all the edges but keep the priors of the top explaining variables while they evaluate on the extracted subgraphs.

\begin{figure}
\centering
\begin{minipage}{.25\textwidth}
    \includegraphics[width=\textwidth]{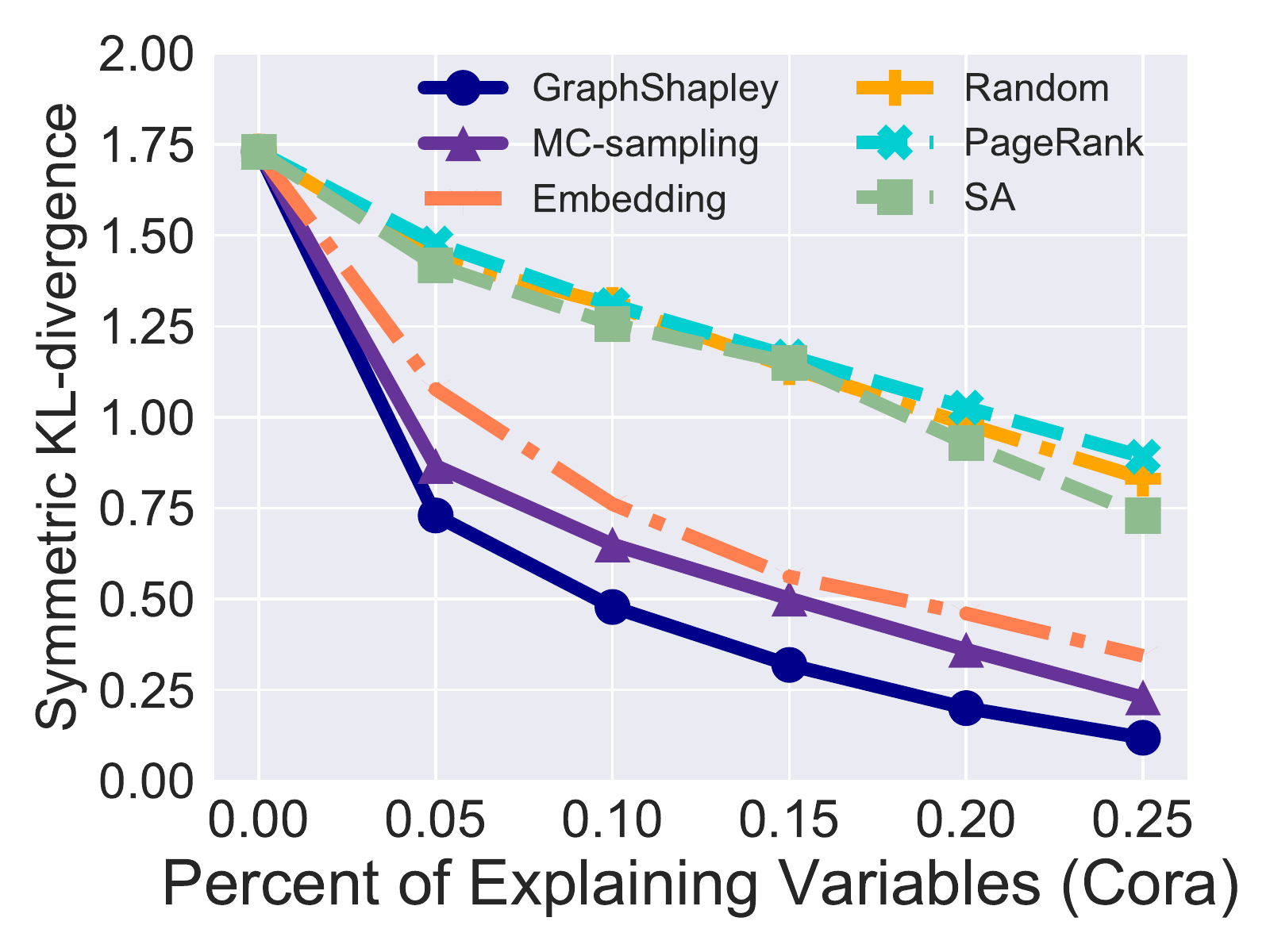}
\end{minipage}%
\begin{minipage}{.25\textwidth}
    \includegraphics[width=\textwidth]{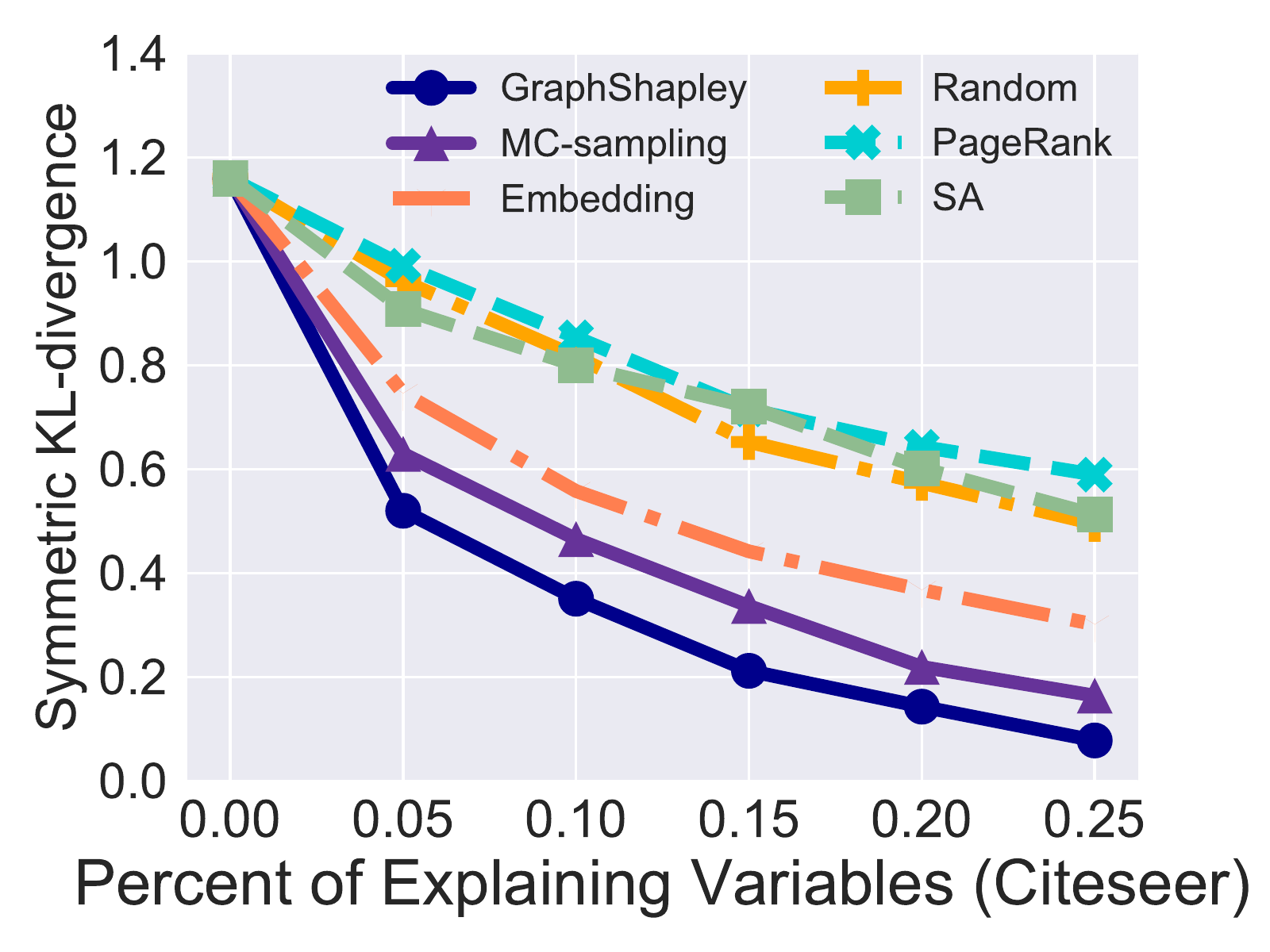}
\end{minipage}%
\caption{
\footnotesize
Symmetric KL-divergence (the smaller the better) on Cora and Citeseer as the percentage of explaining variables increases.
}
\label{fig:KL_change}
\end{figure}

\begin{figure}
\centering
\begin{minipage}{.25\textwidth}
    \includegraphics[width=\textwidth]{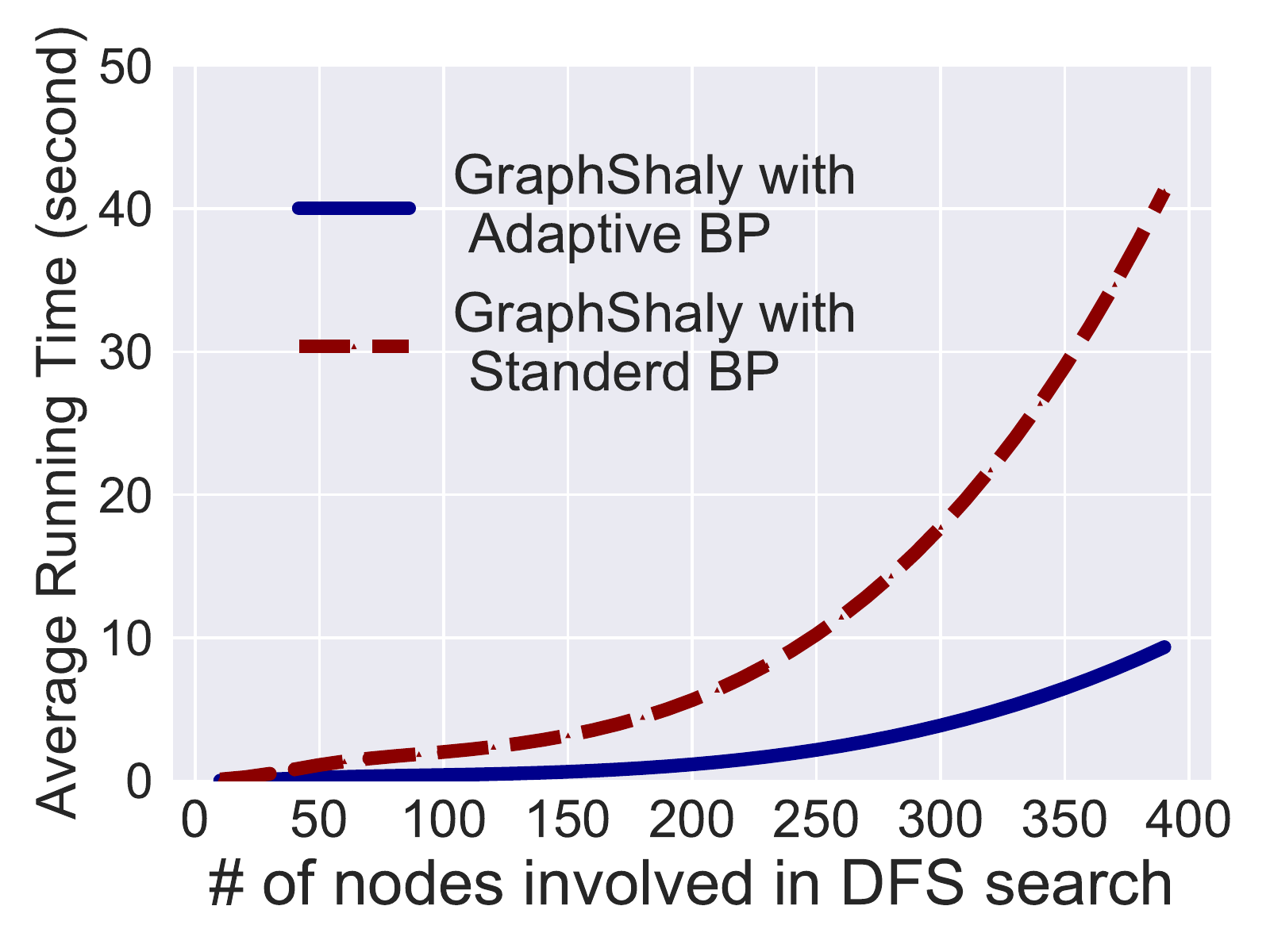}
\end{minipage}%
\begin{minipage}{.25\textwidth}
    \includegraphics[width=\textwidth]{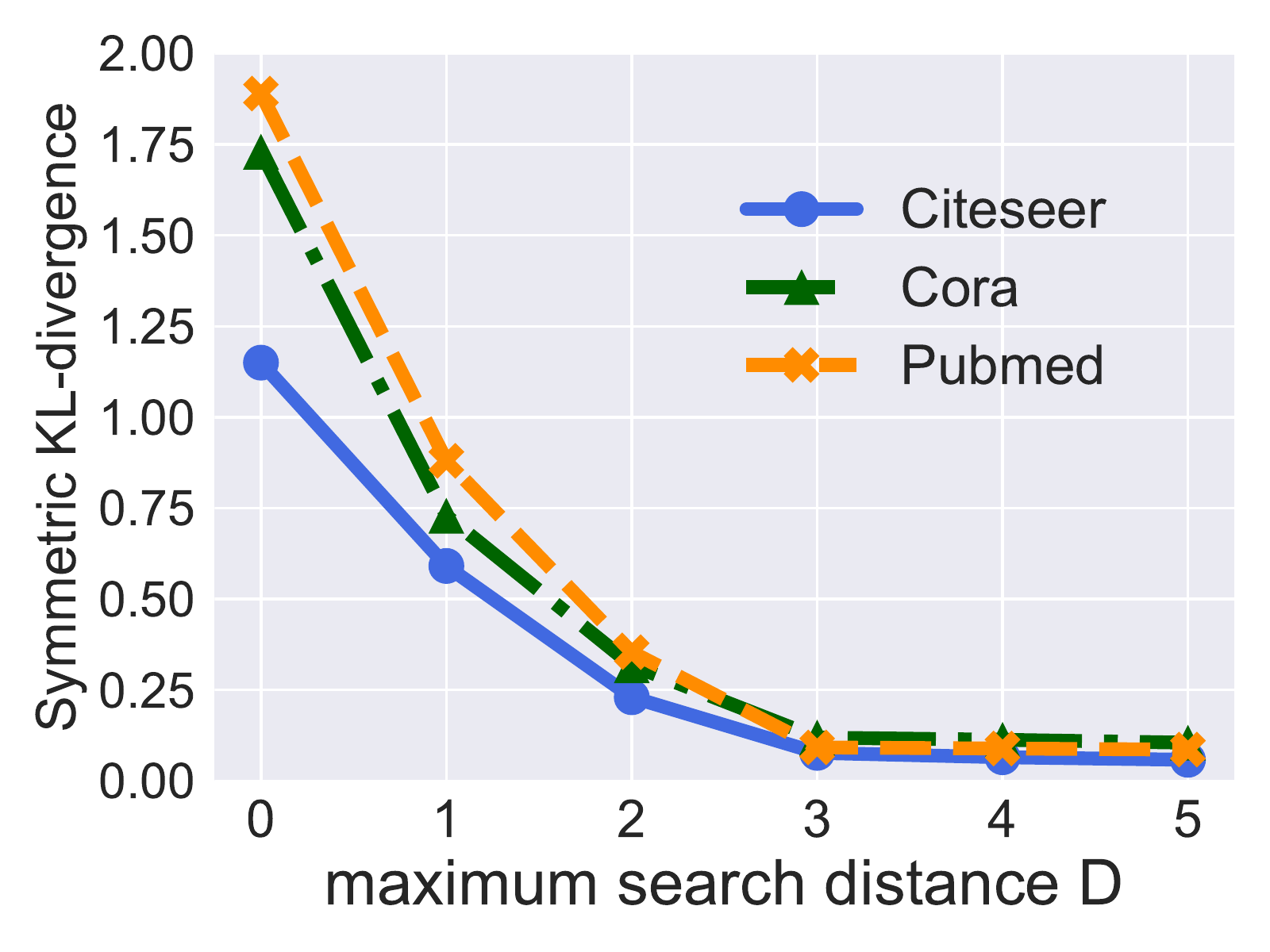}
\end{minipage}%
\caption{\footnotesize
\textbf{Left}: Speed-up by Adaptive BP.
\textbf{Right}: Parameters sensitivity of maximum search distance $D$.
}
\label{fig:speed_up}
\end{figure}

\subsection{Overall Performances}

Unlike classification tasks,
MRF explanations can not be easily evaluated due to the lack of ground truth variable contributions.
After all, there exist multiple explanations of the same observation~\cite{Ross2017}.
We propose a practical experimental protocol to confirm that GraphShapley can identify the probabilistically and topologically influential explaining variables.
All priors on the given MRF are set to the uniform distribution,
and the priors (class-biased or uniform) of the top 25\% of the explaining variables are put back.
Then we compute the approximated $\tilde{b}(X)$ on the ``masked'' MRF.
If the priors of the top variables that making genuine contributions are put back,
$b(X)$ should be approximated well by $\tilde{b}(X)$, measured by
the symmetric KL-divergence (Eq. (\ref{eq:characteristic_function})).
All baselines except for LIME generate a ranking of explaining variables for node selection whose priors are to be put back.
LIME is trained to approximate $b(X)$.
The mean symmetric KL-divergences are shown in Table \ref{tab:overall}, with $t$-tests conducted between GraphShapley and the runner-ups.

We can conclude that: 
1) GraphShapley performs best overall,
due to the consideration of probabilistic and topological contributions.
2) MC-sampling is frequently the runner-up.
However, due to random sampling,
the calculated Shapley values have high variance.
3) LIME has the worst performance, as it is not designed for graphs and cannot take into account the connections on graphical models.

We also change the percentage of top explaining variables for prior assignments.
Figure \ref{fig:KL_change} shows that as more and more priors are retained, the symmetric KL-divergences of all methods go down, with GraphShapley having the fastest decrease, indicating that GraphShapley mostly put the more salient variables before the less relevant ones.

\vspace{.05in}
\noindent\textbf{Speed-up by Adaptive BP.}
We compare the average running time without and with adaptive BP.
Figure \ref{fig:speed_up} (Left panel) plots the running time of the two approaches. It can clearly be seen that running BP from scratches increases the running time exponentially as the number of variables searched by the DFS enumeration algorithm.
On the other hand, Adaptive BP only leads to a near-linear increases in running time.
 
\vspace{.05in}
\noindent\textbf{Parameters Sensitivity}
The maximum distance of the explaining variables $D$ affects the approximation of the Shapley values, with $D=\infty$ leading to the most accurate Shapley values.
On the other hand, a larger $D$ can increase the subgraph search space exponentially.
Figure \ref{fig:speed_up} shows the explanation faithfulness against different $D$ values.

\begin{figure}
    \centering
    \includegraphics[width=0.5\textwidth]{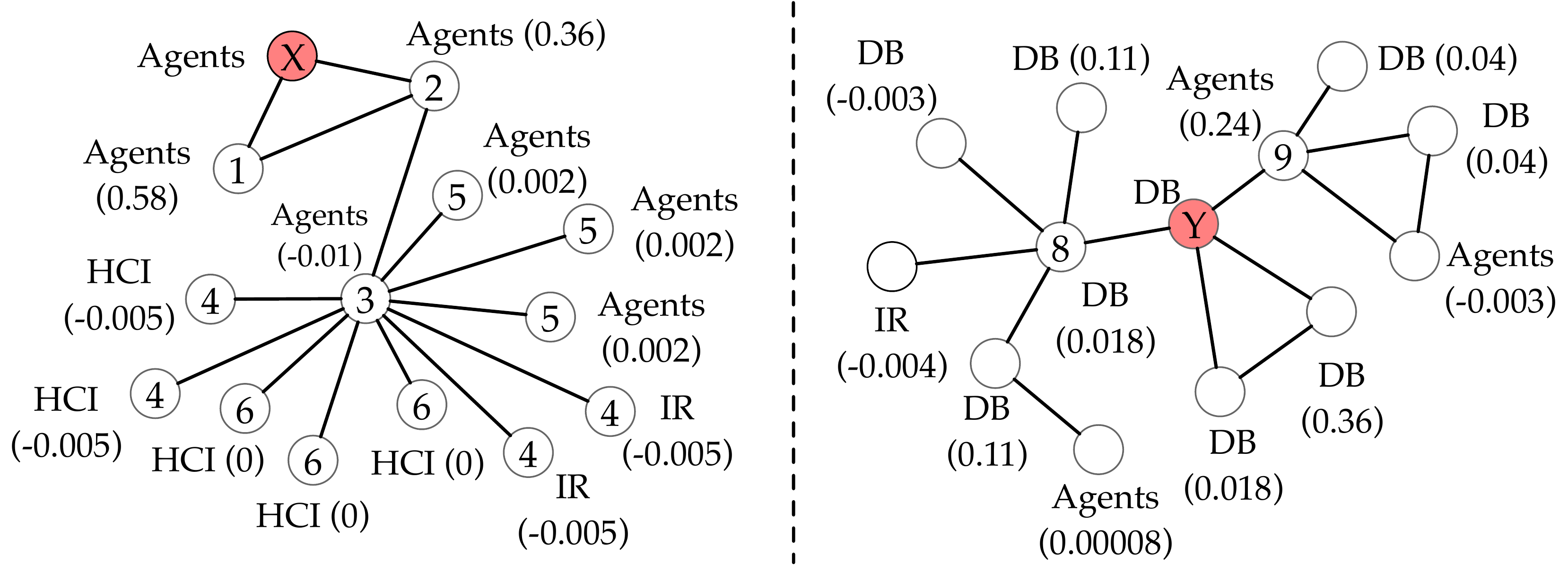}
    \caption{Blank node are explaining variables and the red solid nodes are target nodes.
    The ground truth labels and the computed Shapley values are attached to each explaining node.
    \textbf{Left}: the nodes 1 and 2 have the same labels as node $X$ and are direct neighbor of $X$, thus they have positive higher Shapley values.
    Node 3 has the same label but negative Shapley value, since many of its connected neighbors are of classes ``\textit{IR}'' and ``\textit{HCI}''.
    Four nodes labeled as 4 have equal Shapley values, and the two nodes labeled as 5 have equal Shapley values, verifying the \textbf{equal contribution} proof. Nodes 4 and 6 have different contributions due their different priors.
    Nodes 6 are unlabeled nodes and have no contribution.
    \textbf{Right}: node 9 has positive contribution to the target node, through from a different class. The reason is that the node serves as a bridge transporting the ``\textit{DB}'' probabilities from distance to the target node.
    }
    \label{fig:case}
\end{figure}

\vspace{.05in}
\noindent\textbf{Case Study}
We empirically verify that GraphShapley can capture the probabilistic and topological contribution to the target beliefs and satisfies the \textbf{Equal contribution} property (Theorem~\ref{thm:equal}).
In Figure~\ref{fig:case}, we extract two subgraphs from the MRF for Citeseer 
and show the Shapley values on the nodes.
For the details, see the caption.
The general conclusion is that the probabilistic contributions due to variable prior distribution and the topological contributions due to connectivities can both be captured by GraphShapley.

\section{Related Work}
\textbf{Interpretability and Explanation of Models}
\cite{Ribeiro2016LIME} explains the predictions of a classifier by approximating it locally with an interpretable model.
In \cite{Shrikumar2017DEEPLIFT}
they provide a prediction explanation framework based on Shapley values which encompasses LIME as a special case. 
\cite{ying2019gnn} explains arbitrary graph neural networks using gradients but do not consider PGMs.
\cite{yoon2018inference} uses graph neural networks to learn the message-passing process in belief propagation while explanations for target nodes are not provided.
The more relevant methods are proposed in 
\cite{darwiche2003differential,chan2005sensitivity,chen2019scalable}, which use gradients or greedy subgraph search to find salient variables or subgraphs to explain the inference.
The prominent feature of GraphShapley is that it can quantify both probabilistic and topological contributions.
\noindent\textbf{Shapley Values as explanations}
Shapley values have been applied to the interpretability of machine learning models, but not for PGMs.
Efficient calculation methods of Shapley value have been studied, 
such as in~\cite{Michalak2013efficient,Jia2019efficientshapley}.
Two algorithms with linear complexity for feature importance scoring are developed in \cite{chen2018L-shapley}.
In \cite{Ghorbani2019DATAshapley} and \cite{Ancona2019DNNshapley},
they approximate Shapley values for deep networks via sampling.

\noindent\textbf{Subgraph Enumeration}
Subgraph enumeration algorithms have been researched for multiple decades \cite{yan2002gspan}. 
The most relevant one is proposed in \cite{Skibski2019} where they essentially enumerate all subset of vertices that constitute a connected subgraph for computing Shapley values for a graph-restricted game rather than for explaining BP.
However, in an MRF, the same subset of nodes can be connected in different ways and our enumeration algorithm can address the enumeration of different ways of connection.
\section{Conclusion}
We propose GraphShaley to provide a novel form of explanations for graphical model inference.
The probabilistic and topological contributions of explaining variables can be measured by GraphShapley.
Theoretically, we prove three important theorems to characterize the Shapley values obtained by the algorithm.
In terms of explanation faithfulness and speed,
we empirically show the superior performance of the GraphShapley over other baselines, such as the gradient-based explanations.

\bibliographystyle{named}
\bibliography{ijcai20}
\pdfoutput=1

\end{document}